\documentclass[epj]{svjour}
\usepackage{graphics}
\usepackage[english]{babel}
\usepackage[T1]{fontenc}
\usepackage{amsmath,amssymb}
\usepackage{graphicx}
\begin{document}
\title{Rough infection fronts in a random medium}

\author{A. B. Kolton\inst{1,2,3} \and K. Laneri\inst{1,2}
}                     
\offprints{}          
\institute{Consejo Nacional de Investigaciones Cient\'ificas y T\'ecnicas. \and Centro At\'omico Bariloche (CNEA), R8402AGP Bariloche, Argentina. \and Instituto Balseiro, Universidad Nacional de Cuyo, Argentina.}
%
\date{Received: date / Revised version: date}
%
\abstract{
We study extended
infection fronts advancing over a spatially uniform susceptible population 
by solving numerically a diffusive Kermack McKendrick SIR model with a dichotomous spatially random transmission rate, in two dimensions. 
We find a non-trivial dynamic critical behavior in the mean velocity, in the shape, and in the rough geometry of the displacement field 
of the infective front as the disorder approaches a threshold value for spatial spreading of the infection. 
} 

\PACS{
      {PACS-key}{discribing text of that key}   \and
      {PACS-key}{discribing text of that key}
     } 
\maketitle
\section{Introduction}
\label{intro}
The representation of population heterogeneity in spatially explicit epidemic models was listed as one of the most important challenges \cite{Riley2015}. Very recent results show that spatial transmission variability is essential to reproduce spatio-temporal propagation patterns emerging from some epidemic data sets \cite{RomeoAznar2018}.
One of the simplest epidemic models for infectious diseases is the deterministic Susceptible - Infected - Recovered (SIR) originally formulated by Kermack and McKendrick \cite{Kermack1927} in which the individuals are removed from the population either because they die or because they acquire lifelong immunity. Some infectious diseases affecting humans, like influenza, chickenpox, rabies or rubella, can be modeled using that formulation \cite{RohaniBook}.
Since long time ago, propagation of waves have been observed for several infectious diseases and some of them have been successfully modeled using reaction-diffusion equations. Some examples are the seminal work on plague propagation \cite{Noble1974}, the spatial spread of rabies \cite{MurrayIIBook,Murray86}, Lyme disease \cite{Caraco2002} or Hantavirus \cite{Abramson2003} infection waves.

For all of those natural systems, the substrate in which propagation takes place could be heterogeneous in a variety of ways. For instance, a position dependent transmission would be appropriate for other ecological systems as well, from host specific foliar pathogens \cite{LovettBook} and bacterial colonies \cite{Bonachela2011} to forest fires \cite{Roy2014}. Another approaches account for an spatial heterogeneity on the recovery rate \cite{Capala17} or in the initial distribution of susceptibles \cite{Murray86}. We will here explore another way of introducing spatial heterogeneity that consists on a quenched disordered transmission, that might significantly alter the properties of the propagation front. At variance with most previous approaches, we will consider here a simple disorder with well defined statistical properties. In the same spirit of the general study of diffusion in random media \cite{havlin1987,bouchaud1990} or more specifically interface motion in random media~\cite{BarabasiBook,Kardar1998,Fisher1998} the aim of this approach is to identify the emerging universal statistical features in the transport and geometry of infection waves, those that are independent of the specific realization of the heterogeneity, and other model details.  
In this respect it is worth noting that this approach has been particularly useful for studying interface motion in condensed matter systems~\cite{Kardar1998,Fisher1998}, notably in the case of domain wall motion in ferromagnetic materials \cite{Ferre2013,Ferrero2013}, where quantitative numerical and analytic predictions obtained from solving minimalist models are successfully confirmed experimentally in a remarkably large family of microscopically different systems.

Roughness and velocity of the front, as well as other universal statistical properties related with front propagation in biological systems were measured for example in bacterial colonies cultures with an homogeneous nutrient substrate \cite{Bonachela2011}.  
In the field of epidemiology, the connection between the geometry and the transport of an extended wave has not been addressed, although effectively resembles the dynamics of a growing surface. The formation, structure and dynamics of infection waves can be of course influenced by a big number of factors. Here we will focus in the statistical analysis of non-equilibrium fronts described by the paradigmatic diffusive SIR model.

Specifically, in this paper we study, by numerical simulations, the properties of the propagation front produced by a diffusive SIR model in two dimensions with an heterogeneous random transmission rate. 
It is organized as follows: we start in Sec.~\ref{sec:model} describing the model, the properties of interest, and discussing its behavior qualitatively. In Sec.~\ref{sec:homogeneous} we review some known results for the spatially homogeneous transmission case which are relevant for discussing the inhomogeneous case, analyzed quantitatively in Sec.~\ref{sec:inhomogeneous}. Further discussions and perspectives 
are in the conclusions of Sec.~\ref{sec:conclusions}.

\section{The model and its phenomenology}
\label{sec:model}
We model the coarse-grained dynamics of a local fraction $S({\bf r},t)$ of susceptible individuals and a fraction $I({\bf r},t)$ of infected individuals in a  two dimensional random medium. We assume that the susceptible individuals are immobile and do not die. The susceptible fraction at the position ${\bf r}$ can be converted into infected by local contact with the infected population at a position dependent rate $\beta_{\bf r}$. Infected individuals are considered diffusive with a diffusion constant $D$, they can not recover, and die with an homogeneous death rate $\gamma$.
Under those assumptions, the dynamics of $S$ and $I$ is described by the well known diffusive SIR 
model,~\cite{MurrayIIBook,RohaniBook}:
\begin{eqnarray}
\frac{dS}{dt} &=& 
-\beta_{\bf r} S I \\
\frac{dI}{dt} &=& 
\beta_{\bf r} S I -\gamma I + D \nabla^2 I, 
\label{eq:sir}
\end{eqnarray}
with the recovered or dead fraction not playing any role in the wave dynamics.
We will consider a statistically homogeneous random heterogeneity described by a simple dichotomous noise with probability distribution: 
\begin{equation}
f(\beta_{\bf r})=p\delta(\beta_{\bf r})+(1-p)\delta(\beta_{\bf r}-\beta)
\end{equation}
with $0 \leq p\leq 1$.
In other words, $p$ measures the fraction of space 
where infection can not take place, and can be thus be thought as a randomly "vaccinated" population fraction.
For simplicity, the disorder will be taken isotropic, 
and spatially uncorrelated, such that:
\begin{eqnarray}
\overline{\beta_{\bf r}}&=&(1-p)\beta \label{eq:meanbeta} \\
\overline{\beta_{\bf r}\beta_{\bf r'}}
-\overline{\beta_{\bf r}}^2 &=&\beta^2 p(1-p) \delta({\bf r}-{\bf r'}).
\label{eq:correlationbeta}
\end{eqnarray}
Note also that the disorder is completely characterized by the single parameter $p$, and that the particular $p=0$ case corresponds to the homogeneous case, $\beta_{\bf r}=\beta$.
This quenched disorder thus completely protects the susceptible fraction from infection at the random positions where $\beta_{\bf r}=0$ (occurring with probability $p$), 
but do not alter the diffusive behaviour of infective individuals at those points.

We will be interested in the infection front that is formed by introducing a flat initial infective fraction $I({\bf r}, t=0)=I_0 \theta(\delta x - x)$ on a strip of size $\delta x$ around $x=0$, into an uniform initial susceptible fraction $S({\bf r}, t=0)=S_0$.  We will consider a square medium of size $L \times L$ with Dirichlet boundary conditions in the $x$-direction, $I(x=0,y,t)=S(x=0,y,t)=I(x=L,y,t)=S(x=L,y,t)=0$, and periodic boundary conditions in the $y$-direction, $I(x,y=0,t)=I(x,y=L,t)=0$. The chosen initial and boundary conditions allow us to obtain a unique front propagating in the positive $x$-direction which is flat \textit{on average}. This is quite convenient for the statistical analysis ~\footnote{In general, if the susceptible fraction $S_0$ is large enough, any initial infective fraction produces a large extended infective front at large times (see movies with different initial conditions in the Supplementary information). Since we are interested in the statistical properties of finite segments of the front smooth curvature effects can be neglected. It is thus more convenient to start directly with a flat infective fraction on one side of the sample. This warrants a front that is flat on average, even in presence of the disorder. A well defined statistical analysis of the front fluctuations can be then performed by comparing it with a perfectly flat reference.}.
Equations~\ref{eq:sir} can be easily solved numerically using a finite-difference scheme on a regular lattice (see details of the numerical implementation in the Appendix~\ref{sec:numerics}).

\begin{figure}
\resizebox{0.5\textwidth}{!}{%
  \includegraphics{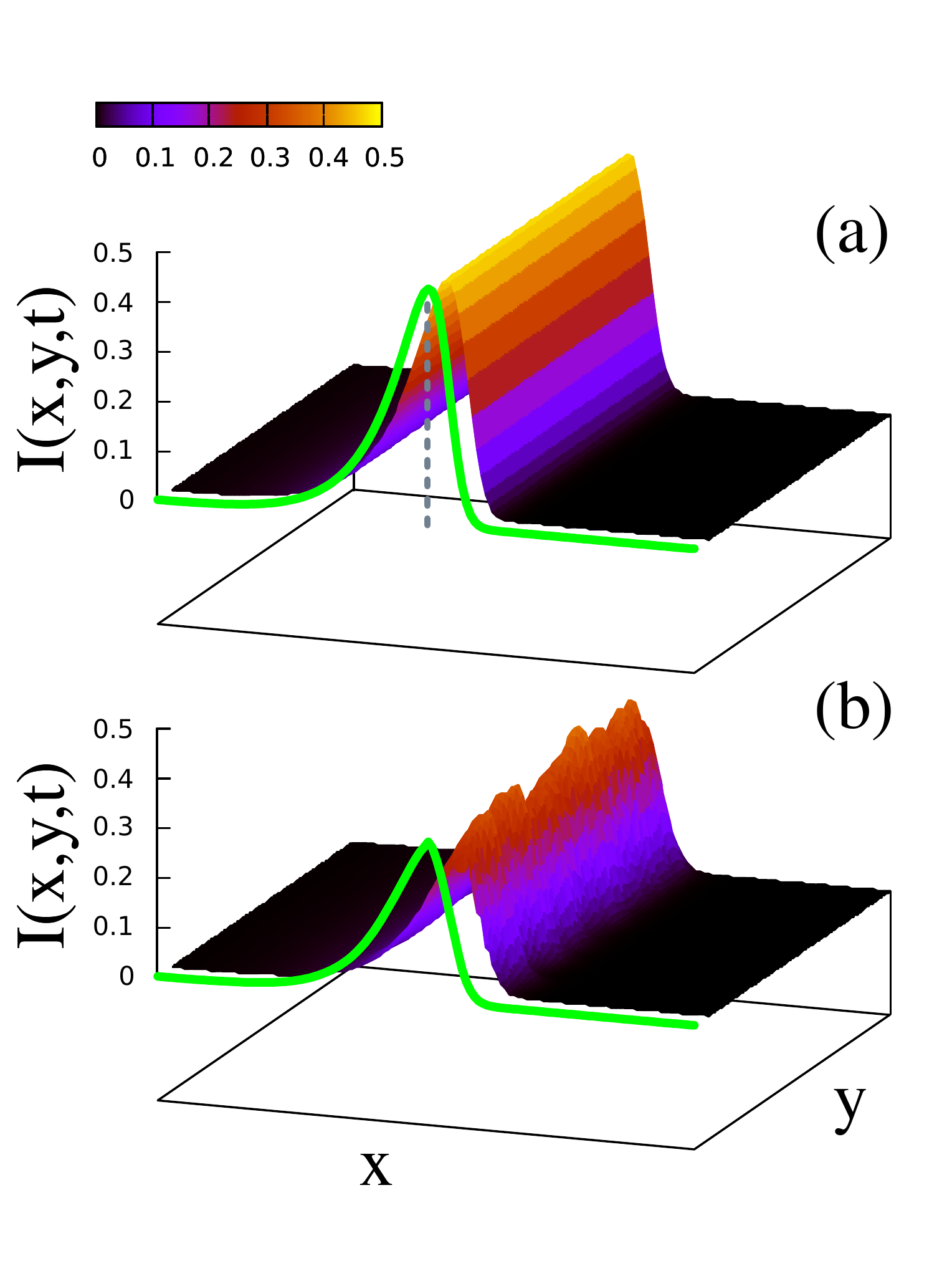}
}
\caption{(color online) Steady-state infective front propagating in the positive $x$-direction (from left to  right), into a large enough and uniform population of susceptibles. The transmission rate is homogeneous ($p=0$) in (a), and random heterogeneous ($p=0.2$) in (b). The color scale indicates the infective fraction. The continuous lines show the average centered front shape, $f_I(x)$ (Eq.~\ref{eq:fIdef}), measured from the wave peak, and the dashed gray line indicates the definition of the average amplitude $I_{\tt max}$ (Eq.~\ref{eq:imaxvst}) of the wave, respectively.
}
\label{fig:surface}       
\end{figure}

For the homogeneous case, corresponding to $p=0$, it is well known (see Appendix \ref{sec:analytics}) that if 
\begin{equation}
S_0 > S_c \equiv \gamma/\beta    
\end{equation}
any $I_0>0$ will trigger a traveling wave, leaving behind a reduced fraction of susceptibles $S_1 < S_c < S_0$.  
After a transient, a steady-state is reached with a flat wave traveling in the $x$ direction, as shown in Fig.~\ref{fig:surface}~(a). A similar traveling wave is also observed at moderate ($p>0$) disorder, as shown in Fig.~\ref{fig:surface}~(b). The steady-state average profile is in general asymmetric, and characterized by a "trailing" and a "leading" edges. The front in presence of disorder presents however some important qualitative differences with respect to the one for $p=0$. We quantify those differences using some statistical observables, that we define in the following paragraphs. 



To characterize the temporal and spatial fluctuations of the front we will be interested in the displacement field of the front $u(y,t)$, defined such that
\begin{equation}
\label{eq:udef}
\max_x[I(x,y,t)]=I(u(y,t),y,t),
\end{equation}
with $u(y,t)$ the $x$-coordinate of the maximum fraction of infected individuals as a function of the coordinate $y$. 
The center of mass position $u_{cm}(t)$ is the spatial average of $u(y,t)$:
\begin{equation}
\label{eq:ucmvst}
u_{cm}(t) \equiv \langle u(y,t) \rangle_y
\end{equation}
where $\langle \dots \rangle_y$ denotes average over the $y$-coordinates.
The infective wave amplitude is thus given by: 
\begin{equation}
\label{eq:imaxvst}
I_{\tt max}(t)=\langle{I(u(y,t),y,t)}\rangle_y.
\end{equation}
where $I_{max}$ denotes the average of the maximum intensity values.
The mean velocity of the front is defined as 
\begin{equation}
\label{eq:cdef}
c \equiv \overline{{\dot u}_{cm}(t)}, 
\end{equation}
where the over line indicates average over disorder and can be replaced by a temporal average in the moving steady-state \footnote{Since the disorder is totally uncorrelated, the front feels different disorder realizations as it moves. This assures the property of self-averaging.}.
The mean amplitude is then 
\begin{equation}
\label{eq:imaxdef}
I_{\tt max}\equiv \overline{I_{\tt max}(t)}.
\end{equation}
The displacement fluctuations can be characterized by the 
mean roughness:
\begin{equation}
\label{eq:w2def}
w^2 \equiv \overline{[u(y,t)-u_{cm}(t)]^2}, 
\end{equation}
or by the structure factor of the front:
\begin{equation}
\label{eq:sofqdef}
S(q) \equiv \overline{|u(q,t)|^2}, 
\end{equation}
where $u(q,t)$ is the spatial Fourier transform of $u(y,t)$~\cite{BarabasiBook}.
We will also be interested in the front shape in the direction of the displacement, 
\begin{equation}
\label{eq:fIdef}
f_I(x) \equiv \overline{\langle I(x-u(y,t),y,t) \rangle_y}, 
\end{equation}
which describes the infective fraction profile from the trailing to the leading edge of the moving front. 


With the above observables we can now describe the main phenomenological differences between the homogeneous ($p=0$) and the disordered ($p>0$) cases.

In Fig.~\ref{fig:surface}~(a)~-~(b) we compare particular snapshots of the infection wave 
for $p=0$ and for $p=0.2$, with $\gamma/\beta=0.2$ in both cases.
The continuous lines, indicating the corresponding functions $f_I(x)$, show that disorder changes the shape of the front. It reduces its amplitude and increases its width when disorder increases. Disorder also breaks the translational symmetry in the $y$ direction present in the spatially homogeneous $p=0$ case.

\begin{figure}
\resizebox{0.5\textwidth}{!}{%
  \includegraphics{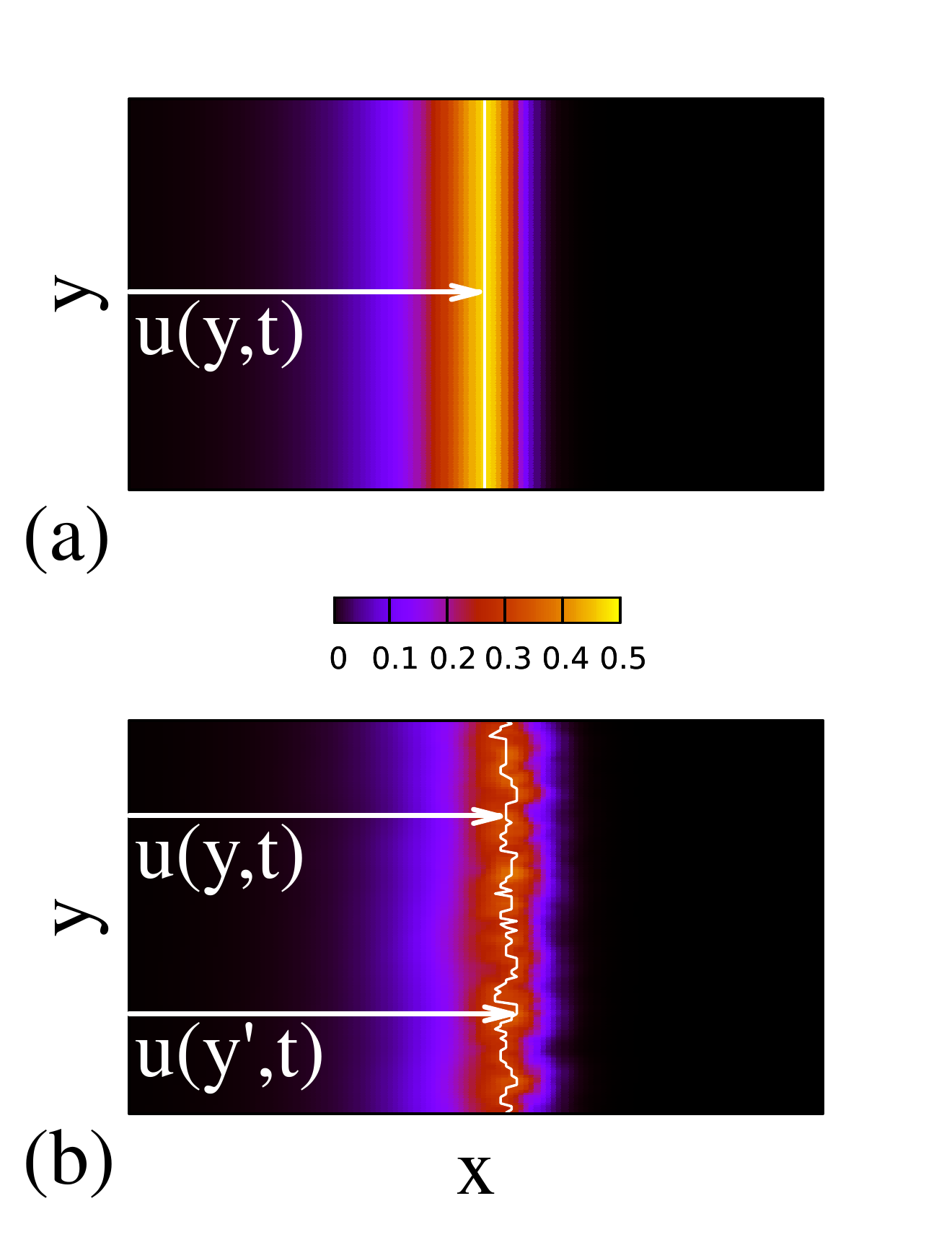}
}
\caption{(color online) 
For the same data as in Fig.~\ref{fig:surface}~(a)~-~(b) we show (white line) the displacement field $u(y,t)$ (Eq.~\ref{eq:udef}) associated to the wave peak as a function of the transverse coordinate $y$. 
The displacement field is flat for $p=0$ and rough for $p>0$. 
}
\label{fig:map}       
\end{figure}

In Fig.~\ref{fig:map}~(a) we show a view map of the infection wave for the spatially homogeneous transmission case for the same parameters of Fig.~\ref{fig:surface}~(a). We indicate the corresponding displacement field $u(y,t)$, defined by Eq.~\ref{eq:udef}. As expected by symmetry considerations, the displacement field is flat and the problem can be reduced to a simpler one dimensional problem, making it more amenable to analytic approaches (see Appendix \ref{sec:analytics}). However $u(y,t)$ is \textit{rough} for $p>0$, as shown in Fig.~\ref{fig:map}~(b).

\begin{figure}
\resizebox{0.5\textwidth}{!}{%
  \includegraphics{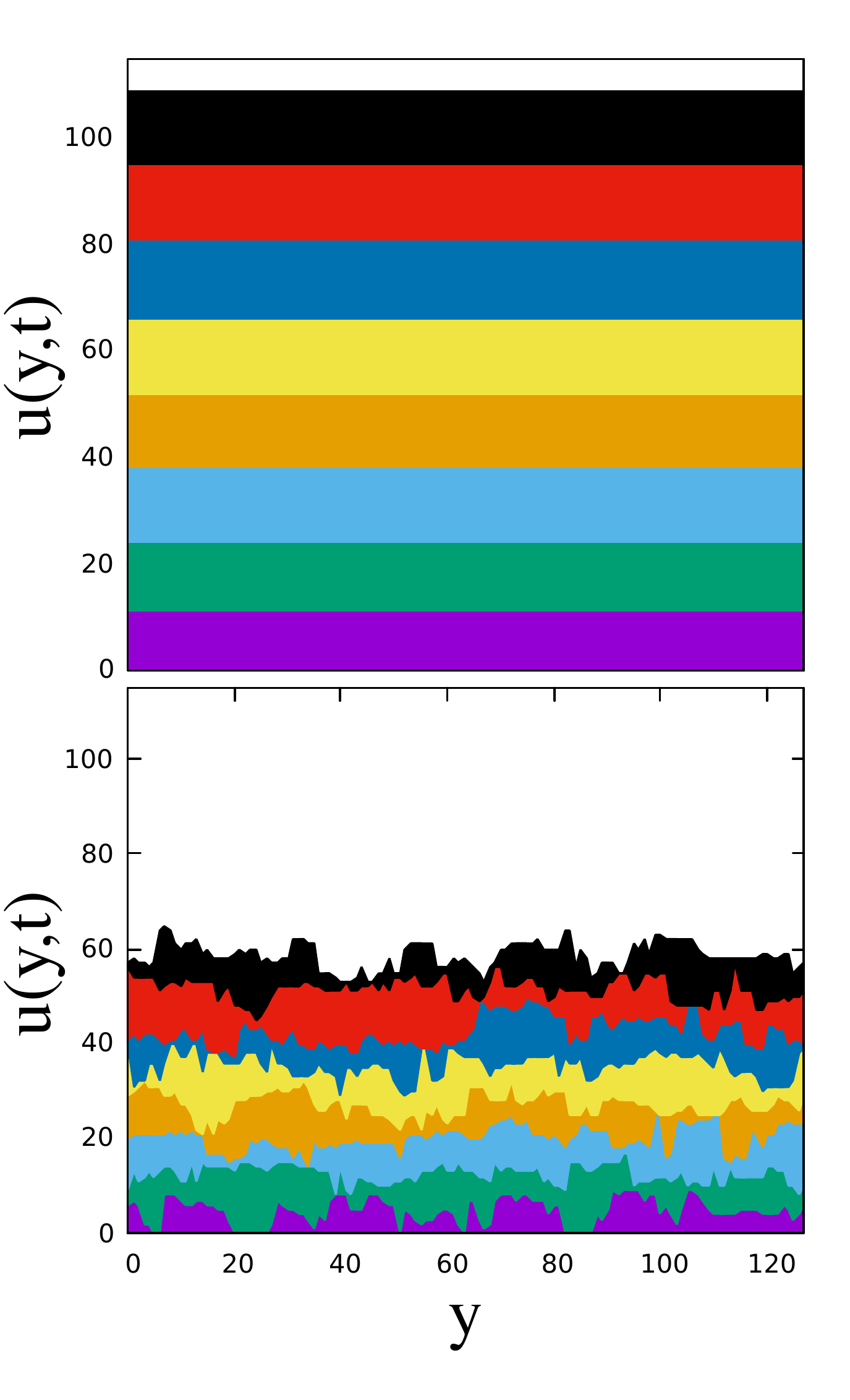}
}
\caption{(color online) 
Temporal sequence of the displacement field $u(y,t)$ of an advancing front for the $p=0$ homogeneous (upper panel) and the $p=0.6$ inhomogeneous (lower panel) cases. Each color or gray level indicates the area spanned in given time interval. The time intervals are the same in both cases showing qualitatively that disorder reduces the mean front velocity.
 }
\label{fig:displacementfield}       
\end{figure}

In Fig.~\ref{fig:displacementfield} we compare the area spanned by the displacement field at regular time intervals in the steady-state, for $p=0$ (upper panel of Fig.~\ref{fig:displacementfield}) and for $p=0.6$ (lower pannel of Fig.~\ref{fig:displacementfield}), with $\gamma$, $\beta$ and $L$ fixed as in the previous figures. Besides the visible spatial roughness of the displacement field for $p=0.6$, we can see that $u(y,t)$ also displays temporal stochastic fluctuations, as the front visits a non repetitive disordered landscape. It is also qualitatively clear that the average speed is reduced by roughly a half for $p=0.6$ with respect to the spatially homogeneous transmission $p=0$ case.

Summarizing, the spatially inhomogeneous transmission rate $\beta_{\bf r}$ introduced in Eq.~\ref{eq:sir}, produces the following qualitative effects with respect to the homogeneous case:
\begin{enumerate}
\item It breaks the translation symmetry of the problem in both directions, $x$ and $y$, producing spatio-temporally fluctuating fronts. The temporal fluctuations and lateral spatial fluctuations of the infective front can be characterized by the rough displacement field $u(y,t)$.
\item It changes the average shape $f_I(x)$ of the front in the direction of its mean displacement, making it wider and reducing its amplitude $I_{\tt max}$. 
\item It reduces the average velocity $c$ of the front.
\end{enumerate}
As we will discuss below, all these effects persist by increasing $p$ up to a well defined critical value $p_c$, near which $c$, $I_{max}$ tend to vanish and,  concomitantly, $w^2$ tend to diverge, all displaying a non-trivial critical behavior. For $p>p_c$ disorder completely stops the propagation even if $S_0>S_c$. The goal of our paper is to find $p_c$ and to quantify the dynamical and geometric properties of the front as a function of $p$, from the homogeneous $p=0$ case to the critical $p \to p_c$ case.



%

\section{Homogeneous case}
\label{sec:homogeneous}

\begin{figure}
\resizebox{0.5\textwidth}{!}{%
  \includegraphics{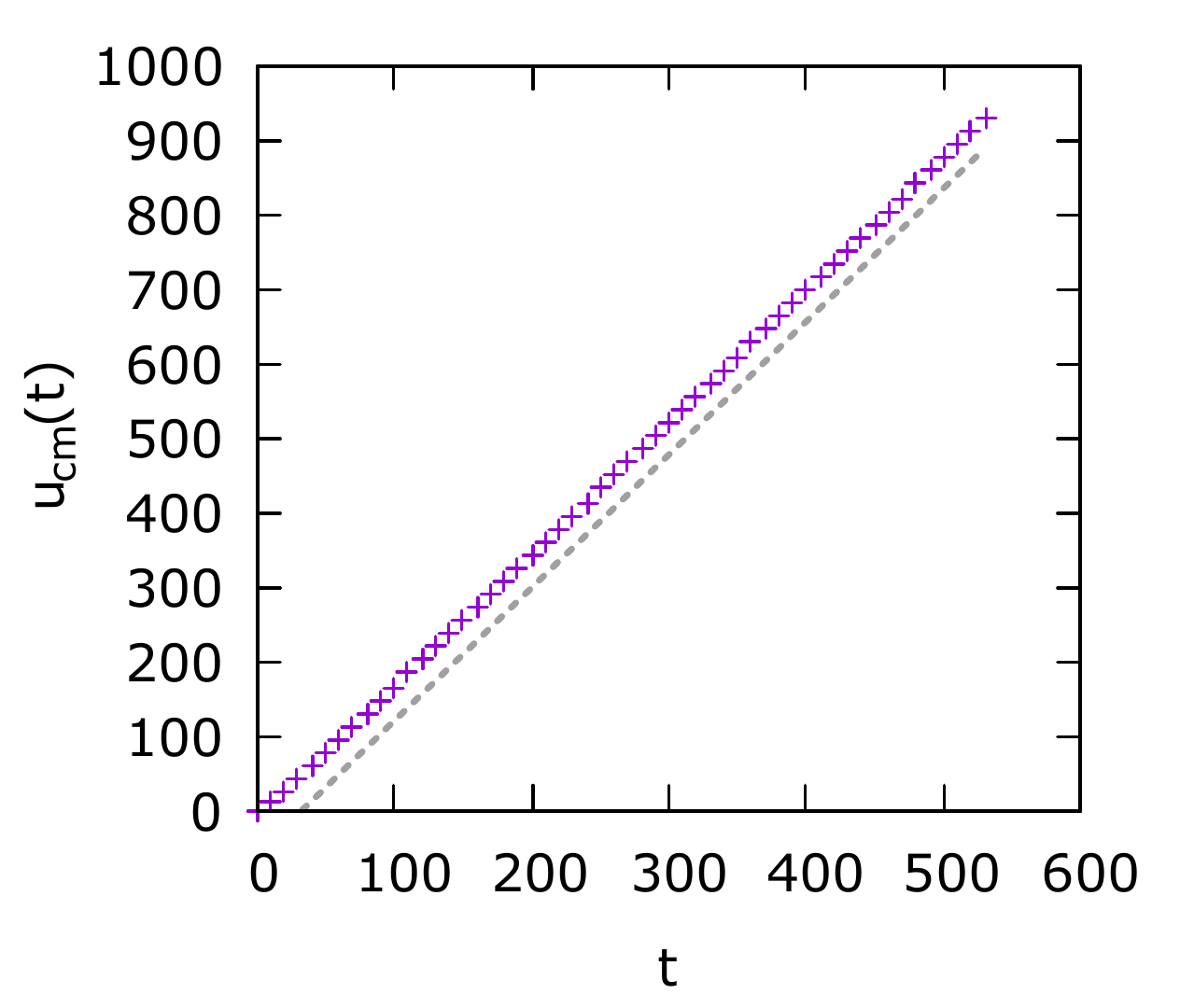}
}
\caption{(color online) 
Center of mass of the displacement field $u_{cm}$ vs time $t$ for the homogeneous case ($p=0$). After a short transient, $u_{cm} \sim c_0 t$.
The dashed line corresponds to the analytic solution of Eq.~\ref{eq:cclean} for $c_0$.
}
\label{fig:uvstclean}       
\end{figure}
\begin{figure}
\resizebox{0.5\textwidth}{!}{%
  \includegraphics{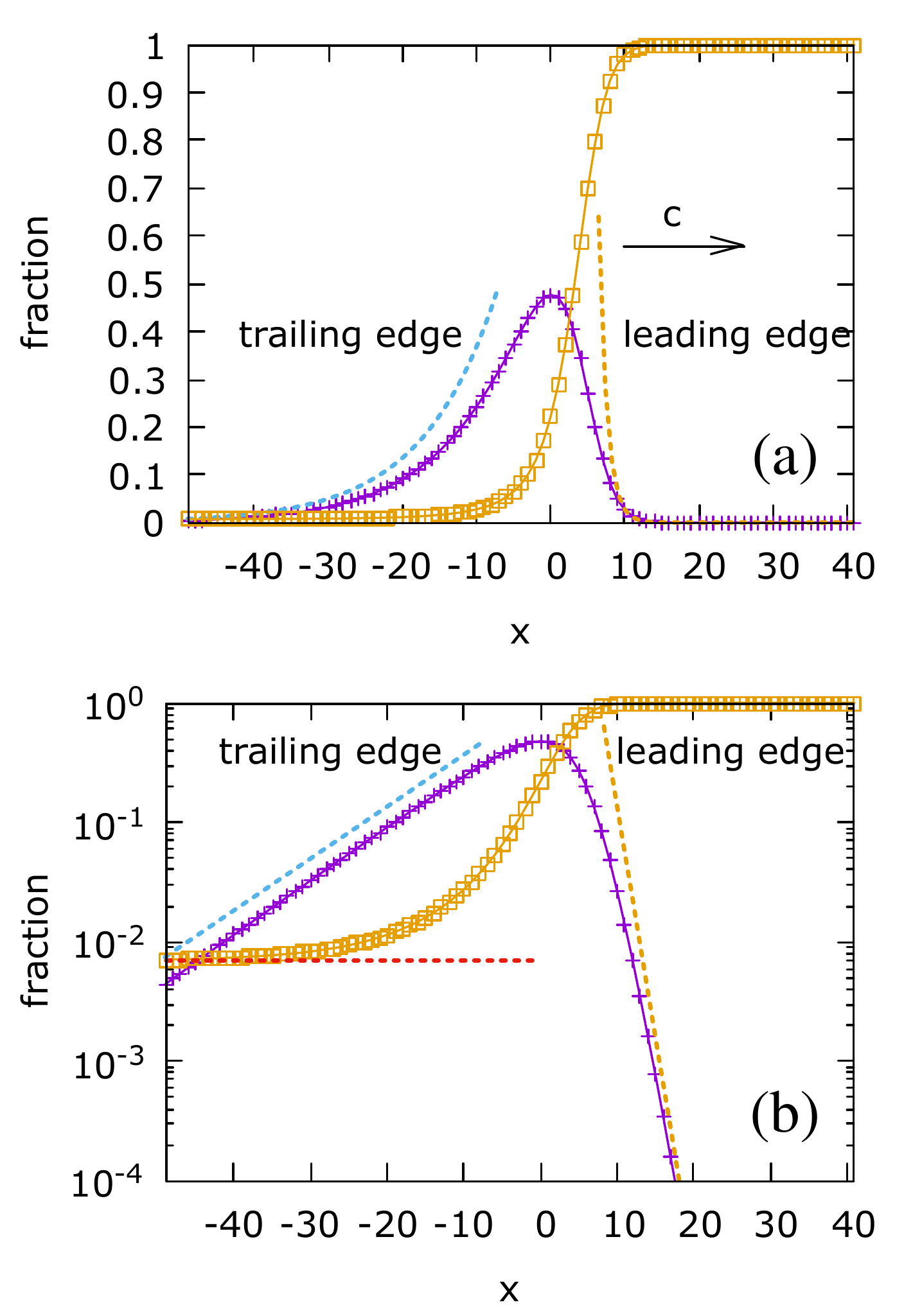}
}
\caption{(color online)
Shape of the infective (crosses) and susceptible (squares) fronts, centered at the maximum infection fraction, in linear (a) and log-linear (b) scales. 
The tilted dashed lines correspond, as inticated, to the analytic asymptotic solutions describing the leading and trailing edges, 
Eqs.~\ref{eq:leading} and \ref{eq:trailing}, respectively. The horizontal dashed line in (b) corresponds to the analytic solution for the fraction of remaining susceptibles $S_1$ after the wave passage Eq.~\ref{eq:s1clean}.
}
\label{fig:cleanedges}       
\end{figure}

Before tackling the disordered case we review the homogeneous case, corresponding to $p=0$. This case has been extensively studied in the past and many properties of its steady-state solution can be obtained analytically (see Appendix \ref{sec:analytics}). We review here the most relevant properties for our study.

For the flat initial condition a steady-state traveling wave solution, of the form $I(x,y,t) \equiv f_I(x-c_0 t,t)$, exists only for $S_0 > S_c \equiv \gamma/\beta$. For the initial and boundary conditions chosen, the traveling front is perfectly flat and invariant with respect to the $y$-axis, and for large times we simply have $u(y,t) \sim c_0 t$, without temporal fluctuations. The homogeneous velocity $c_0$ is given by (see Appendix \ref{sec:analytics}):
\begin{equation}
c_0=2\sqrt{D(\beta S_0-\gamma)}
=2\sqrt{D\beta(S_0-S_c)}.
\label{eq:cclean}
\end{equation}
A steadily moving front is then possible only if the mentioned condition $S_0>S_c$ is met. We also note that the diffusivity and transmission rate both contribute to increase the average speed $c_0$. 
Interestingly, the above expression for $c_0$ is basically determined by what happens in the leading edge of the front, where the system of Eqs.~\ref{eq:sir} can be linearized and hence solved analytically~(see for instance \cite{MurrayIIBook}). In Fig.~\ref{fig:uvstclean} we compare the mean velocity $c\approx 1.79$ corresponding to $\gamma=0.2$, $\beta=1$ and $S_0=1$ with a numerical solution, showing an excellent agreement.

The asymptotic shape of the front can be also obtained analytically ~(see Appendix \ref{sec:analytics}), yielding the right tail or leading edge:
\begin{equation}
f_I(x) \sim \exp \left[-\frac{c_0}{2D} x \right].
\label{eq:leading}
\end{equation}
Interestingly, there is a sort of "Lorentz" contraction of the front: the faster the front moves, the sharper its leading edge is, decaying exponentially to zero in a characteristic distance $D/c_0$.  A similar calculation applies for the asymptotic shape of the trailing edge or left tail of $f_I(x)$, which far enough from the infection peak also decays exponentially but at a slower spatial rate 
\begin{eqnarray}
f_I(x) &\sim& \exp \left[\left(\frac{c_0}{2D} - \sqrt{\left(\frac{c_0}{2D}\right)^2 - \frac{(\beta S_1 - \gamma)}{D}}\right)x \right] \\ &=& \exp \left[\left(\frac{c_0}{2D} - \sqrt{\beta(S_0-S_1)/D}\right)x \right],
\label{eq:trailing}
\end{eqnarray}
where $S_1$ is the fraction of susceptibles left after the passage of the wave, given by the transcendent equation:
\begin{equation}
\frac{S_1/S_0 - 1 }{\ln S_1/S_0}= \frac{\gamma}{\beta S_0}.
\label{eq:s1clean}
\end{equation}
The above equation implies that $0< S_1 < S_c < S_0$, showing that, in the steady-state, the infected fraction in the trailing edge can not trigger a backward moving wave.
In Fig.~\ref{fig:cleanedges}~(a)~-~(b) we compare these predictions (derived in Appendix \ref{sec:analytics}) with numerical results 
(see implementation details in Appendix~\ref{sec:numerics}) for $p=0$. 
The analytic results fairly fit the asymmetric tails of the infective front, and the left tail of the susceptible fraction shows an excellent agreement with $S_1$, as can be appreciated in Fig.~\ref{fig:cleanedges}~(b).

\section{Inhomogenous case}
\label{sec:inhomogeneous}
In the presence of quenched disorder ($p>0$), an analytic calculation of the steady-state statistical properties of Eq.~\ref{eq:sir} becomes difficult. We then solve the equations 
numerically, as explained in Appendix~\ref{sec:numerics}.
\subsection{Steady-state equilibration}
\label{sec:transients}

\begin{figure}
\resizebox{0.5\textwidth}{!}{%
  \includegraphics{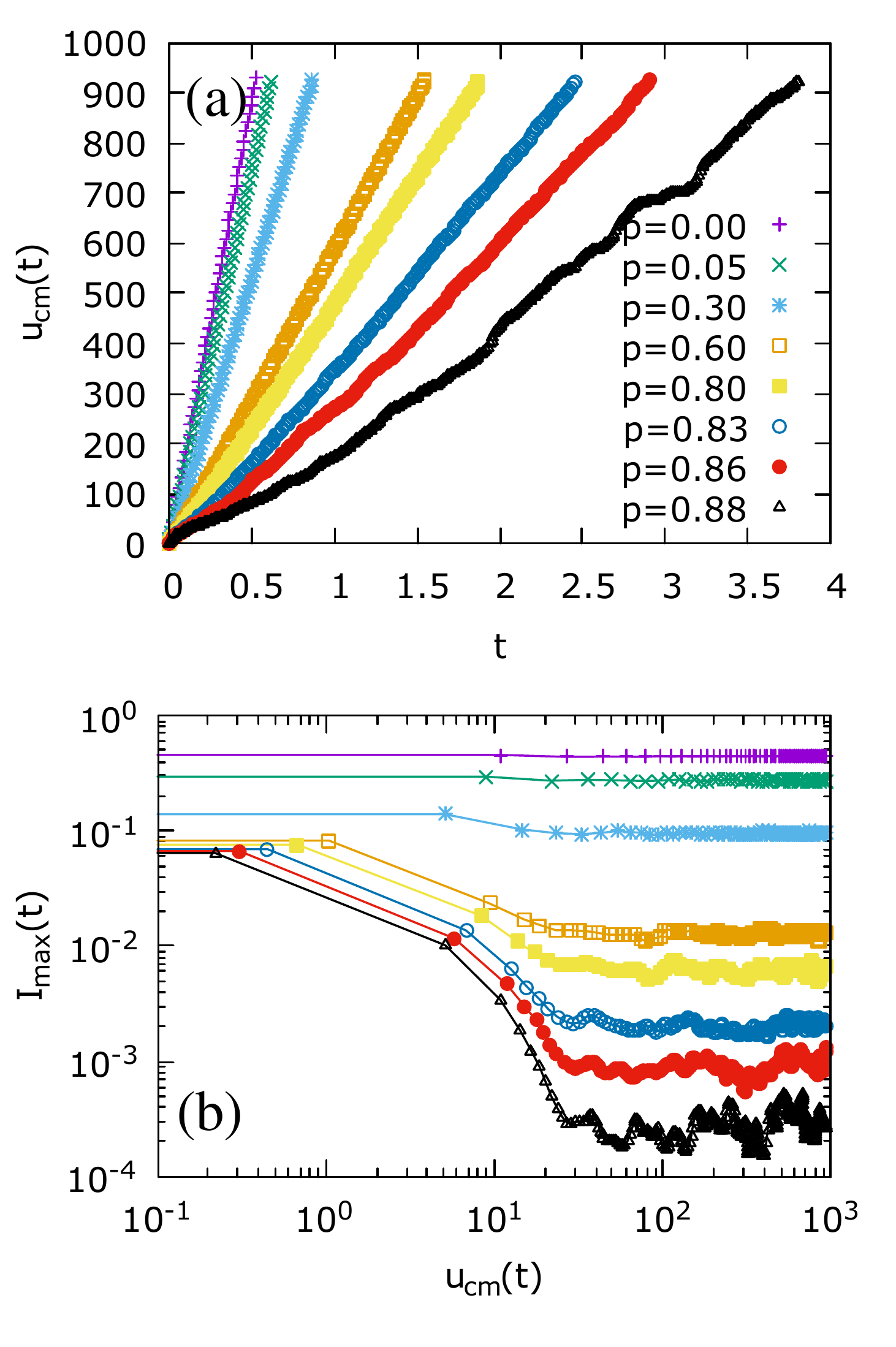}
}
\caption{Steady-state equilibration of the velocity $c$, defined as hte long-time limit $u_{cm} \sim c t$ (a), and of the front amplitude $I_{\tt max}(t)$ (b), for different disorder amplitudes $p$ . In (b) we evidence that the steady-state equilibration time is roughly controlled by a fixed displacement of the front and thus by $c^{-1}$. The data, corresponding to a single realization of disorder, also illustrate the temporal fluctuations induced by disorder in both the transient and steady states in a finite system.
}
\label{fig:transient}       
\end{figure} 
The steady-state equilibration of the system occurs after a transient. 
We find that the mean velocity $c$ and the shape $f_I(x)$ of the front are the faster observables to converge to their steady-state values.
Fig.~\ref{fig:transient}~(a) shows how a steady-state velocity is reached for different values of $p$, corresponding to a linear dependence of $u_{cm}$ with $t$. A linear fit at long times gives an estimate of $c$, for each $p$. Fig.~\ref{fig:transient}~(b) shows how the fluctuating amplitude $I_{\tt max}(t)$ reaches a statistically steady-state after a transient. The data shown, corresponding to a single realization of disorder, also illustrate the temporal fluctuations induced by disorder in both the transient and steady states in a system of size 
$2048 \times 2048$ sites. Since disorder is completely uncorrelated and the rough wave relatively well localized, we find that disorder realization can be replaced by temporal average in the steady-state. 
It is worth noting in Fig.~\ref{fig:transient}(b) that by plotting $I_{\tt max}(t)$ as a function of $u_{cm}$ we see that the transient time roughly corresponds to a fixed value of $u_{cm}$ for all $p$. This suggests that the transient in these global quantities is basically controlled by $\sim c^{-1}$. From Figs.~\ref{fig:transient}~(a)~and~(b) we observe that both the steady-state velocity $c$ and infection amplitude $I_{\tt max}$ decrease and tend to vanish with increasing $p$. Interestingly, since the equilibration time of these quantities grows as $\sim c^{-1}$, the equilibration time tend to diverge with increasing $p$. In the following sections we discuss the steady-state and show that there is a unique critical value $p_c<1$ such that for $p>p_c$ the spreading of the infection stops.

\subsection{Front Velocity}
\label{sec:c}
\begin{figure}
\resizebox{0.5\textwidth}{!}{%
  \includegraphics{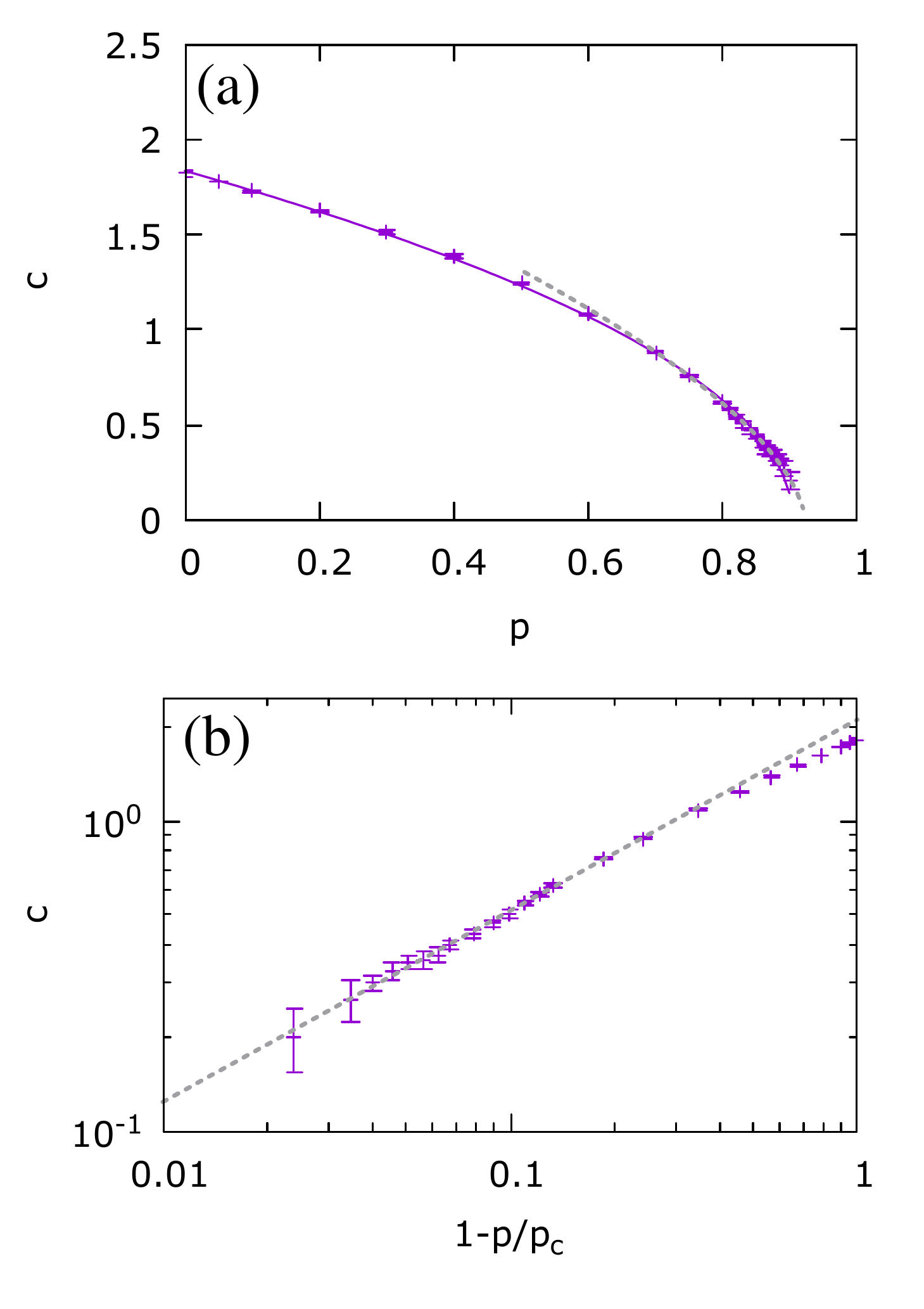}
}
\caption{Mean front velocity as a function of the disorder parameter $p$. (a) The solid line shows a fit to the power law $(1-p/p'_c)^{1/2}$, with $p'_c \approx 0.9$ for the whole range of $p$ where $c$ is positive. The dashed line is the best power law fit for vanishing values of $c$. In (b) we show that the critical behaviour is better described by $(1-p/p_c)^{\alpha_c}$ with $\alpha_c\approx 0.6$ and $p_c=0.92$.}
\label{fig:cvsp}       
\end{figure}

In Fig.~\ref{fig:cvsp}~(a) 
we plot the behaviour of the velocity $c$ vs $p$. As it can be appreciated $c$ tends to vanish at a critical value $p_c$. Interestingly $c \approx  c_0 (1-p/0.9)^{1/2}$ with $c_0$ given by Eq.~\ref{eq:cclean} fairly fitting the whole curve, from $p=0$ to $p \approx 0.9$. 
A closer inspection into the region where $c$ is very small  reveals however that there exists a different, more accurate power-law behaviour $c \approx (1-p/p_c)^{\alpha_c}$, with ${\alpha_c} \approx 0.6 \pm 0.05$ and $p_c \approx 0.92 \pm 0.02$ (see Fig.~\ref{fig:cvsp}~(b)).
This behaviour is reminiscent of continuous phase transitions. 
We can thus think $p$ as the control parameter, $c$ as an order parameter, $p_c$ the critical threshold and $\alpha_c$ the characteristic critical exponent of the transition. Moreover, since the equilibration time goes like $\sim c^{-1}$ (see Section \ref{sec:transients}), it tends to diverge at $p_c$. This is the analogue of the critical slowing down of continuous phase transitions.

It is worth comparing the above results with a naive homogenization approach. It consists in replacing $\beta$ in Eq.~\ref{eq:cclean} by an effective transmission $\beta_{\tt eff}(p)$ assuming it is well approximated by the spatially averaged transmission 
\begin{equation}
\beta_{\tt eff}(p) \equiv \overline{\beta_{\bf r}}=(1-p)\beta    
\label{eq:homogeneization}
\end{equation}
(see Eq.\ref{eq:meanbeta}). We thus obtain 
\begin{equation}
c^{\tt eff}(p) \approx 2\sqrt{D(\beta(1-p) S_0-\gamma)} = c_0 (1-p/p^{\tt eff}_c)^{\alpha^{\tt eff}_c}, 
\label{eq:c_homogeneization}
\end{equation}
predicting critical behaviour at $p^{\tt eff}_c = 1-\gamma/\beta S_0$ with a critical exponent $\alpha^{\tt eff}_c = 1/2$. This critical behaviour is qualitatively similar to what is observed in Fig.~\ref{fig:cvsp}. However, for the parameters used in Fig.~\ref{fig:cvsp} we get $p^{\tt eff}_c=0.8$, different from the $p_c \approx 0.92 \pm 0.02$ obtained from the simulations. In addition, $\alpha_c \approx 0.6 \pm 0.05$, is different from the predicted $\alpha^{\tt eff}_c = 1/2$. In other words, the naive homogenization approach is inaccurate for predicting the critical behaviour of $c(p)$. This is the first indication that the observed critical behaviour may be non-trivial, as it will become even more evident in the next sections.

\subsection{Front Amplitude}
\begin{figure}
\resizebox{0.5\textwidth}{!}{%
  \includegraphics{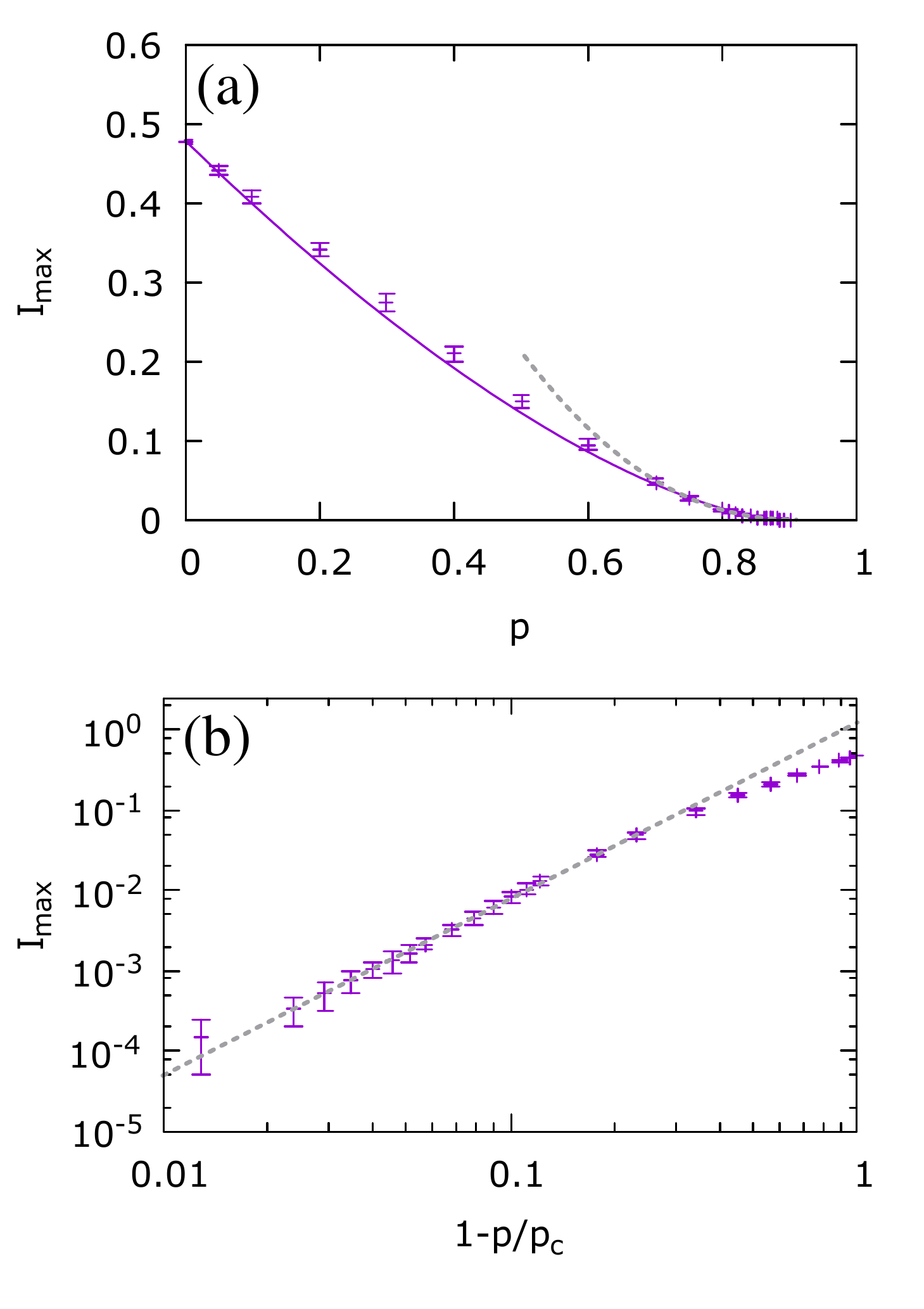}
}
\caption{Average amplitude of the infection wave $I_{max}$ vs disorder parameter $p$. The solid line in (a) shows an overall power-law fit $I_{max}\sim (1-p/0.89)^{1.5}$. The dashed line shows a power-law fit near the critical region, giving 
$I_{max} \approx (1-p/p_c)^{\alpha_{I}}$ 
with $\alpha_{I} \approx 2.2$.
}
\label{fig:Imaxvsp}       
\end{figure}

In Fig.~\ref{fig:Imaxvsp} 
we show the behaviour of 
the front amplitude $I_{\tt max}$
vs $p$. An approximate power-law $I_{max} \approx (1-p/0.89)^{1.5}$ fits the complete range of $p$, as shown 
with a solid line in Fig.~\ref{fig:Imaxvsp}~(a).
For vanishing values of $I_{\tt max}$, however, a more accurate power-law $I_{\tt max} \approx (1-p/p_c)^{\alpha_{I}}$, with $p_c\approx 0.91 \pm 0.02$ and $\alpha_{I} \approx 2.2 \pm 0.05$, is found. This is consistent with the existence of a single critical point at $p_c \approx 0.91 \pm 0.02$, in agreement with the critical behaviour of $c(p)$ shown in Fig.~\ref{fig:cvsp}~(b). 

\subsection{Front shape}
\begin{figure}
\resizebox{0.5\textwidth}{!}
{%
 \includegraphics{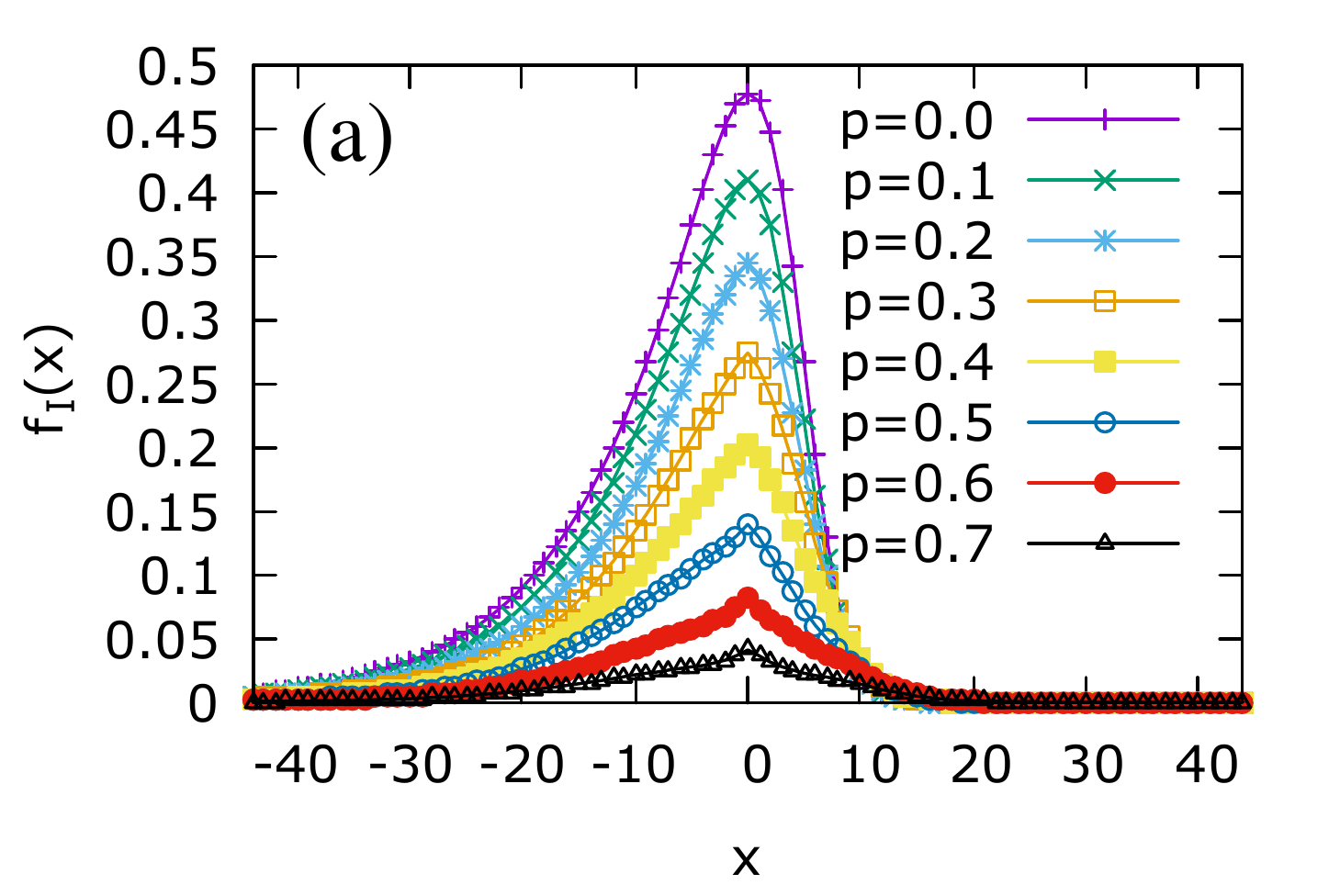}//
}
\resizebox{0.5\textwidth}{!}
{%
 \includegraphics{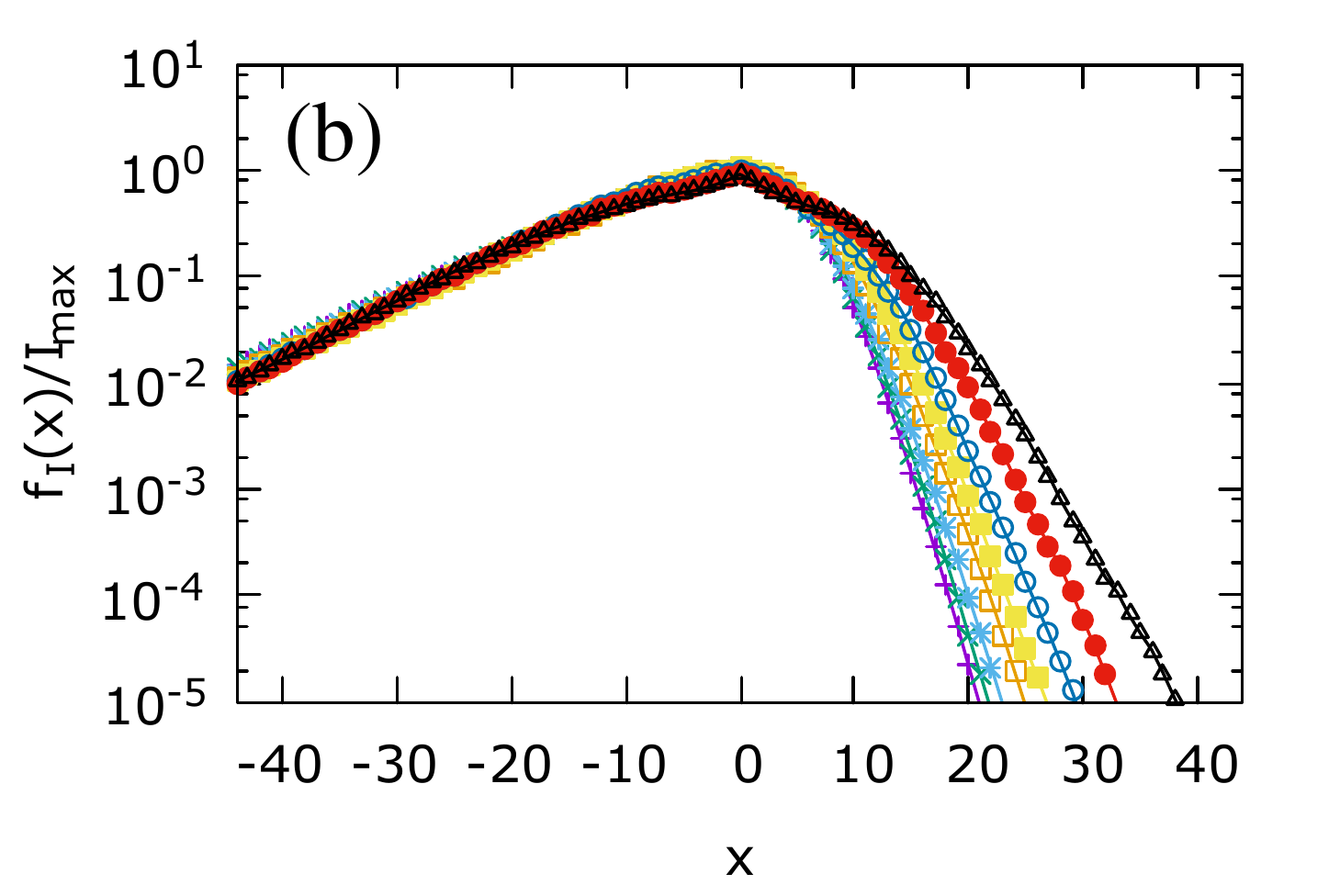}
}
\resizebox{0.5\textwidth}{!}
{%
 \includegraphics{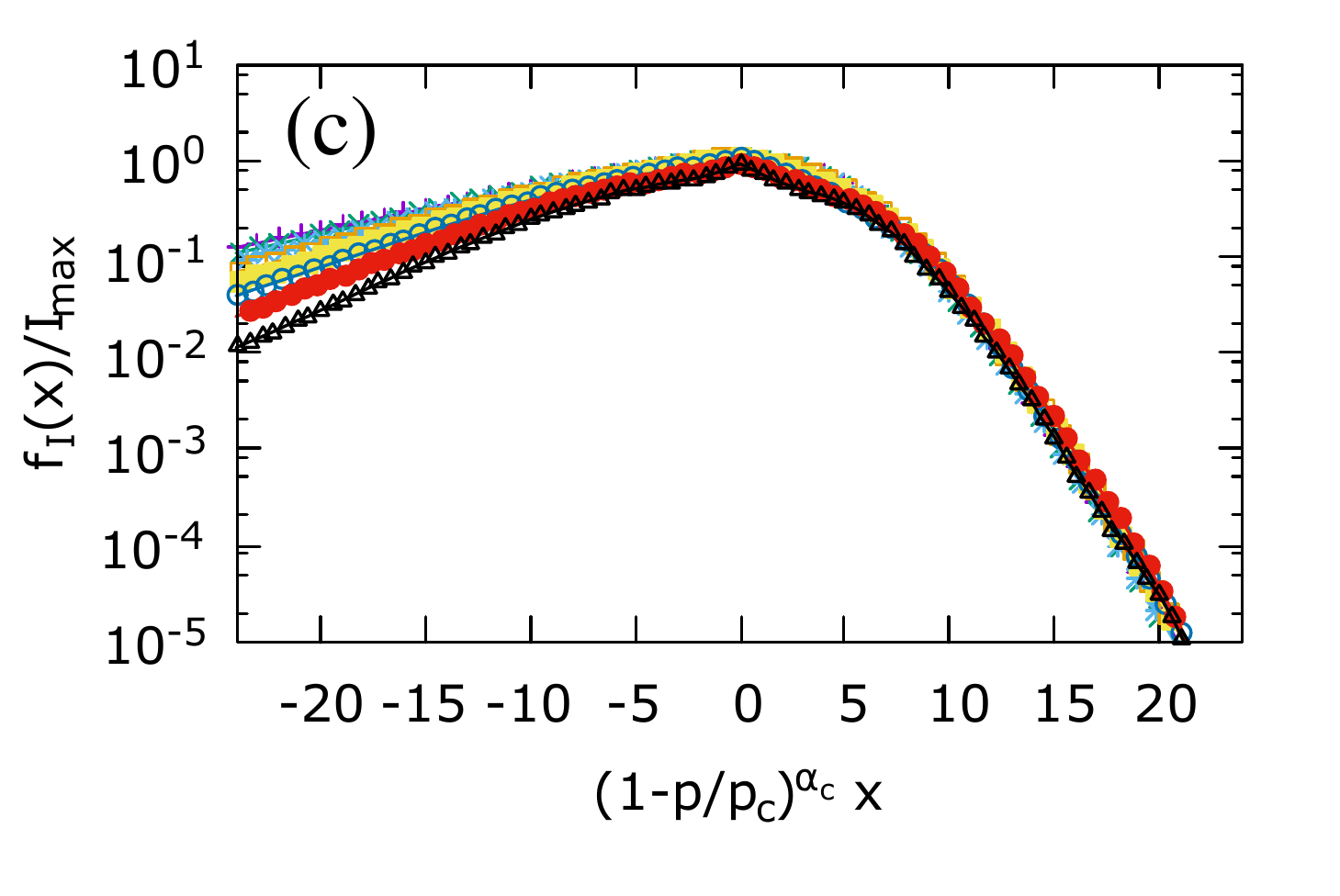}
}
\caption{Front shape function $f_I(x)$ vs the disorder parameter $p$, without rescaling (a), rescaled only by the mean amplitude $I_{\tt max} \equiv f_I(0) \approx (1-p/p_c)^{\alpha_I}$ with $\alpha_I \approx 2.2$ (Fig.~\ref{fig:Imaxvsp}~(b)), and 
additionally rescaled in the $x$ axis by $(1-p/p_c)^{\alpha_f}$ with $\alpha_f \approx 0.4$. }
\label{fig:shape}       
\end{figure}

In Fig.~\ref{fig:shape}~(a) 
we show the evolution of the average front shape as a function of $p$. It can be observed that an increase of $p$ reduces the amplitude of the front, as noticed in the previous section, and also reduces its asymmetry, by making the leading edge less sharp.
Interestingly, in Fig.~\ref{fig:shape}~(b) we show that besides the change of amplitude, the exponential decay rate of the trailing edge remains practically unchanged with respect to the homogeneous $p=0$ case. This result is in sharp contrast with the behaviour of the leading edge, whose exponential decay rate display a critical behaviour, vanishing as 
$(1-p/p_c)^{\alpha_f}$ with $\alpha_f \approx 0.4 \pm 0.05$, 
as evidenced by the re-scaled shape function shown Fig.~\ref{fig:shape}~(c). 
We also note that the shape function $f_I(x)$ develops a curious cusp at its center for large values of $p$.

It is again interesting to compare the above results with the ones predicted by the naive homogenization procedure of Eq.~\ref{eq:homogeneization}. 
If we apply it to Eq.~\ref{eq:leading}, describing the leading edge, we get:
\begin{equation}
    f^{\tt eff}_I(x) \sim \exp \left[-\frac{c^{\tt eff}(p)}{2D} x \right] 
    = \exp \left[-\frac{c_0(1-p/p^{\tt eff}_c)^{1/2}}{2D} x \right].
\end{equation}
While the characteristic spatial decay of the leading edge observed in the 
simulations goes like $\sim (1-p/p_c)^{\alpha_f}$ (see Fig.~ \ref{fig:shape}~(c)) the predicted decay goes as $\sim (1-p/p^{\tt eff}_c)^{1/2}$.
The exponent $\alpha_f = 0.4 \pm 0.05$ is close to the predicted $1/2$ withing the error bars, but the predicted threshold is again clearly below the numerical one.

%

 \subsection{Dynamic roughening}
\begin{figure}
\resizebox{0.5\textwidth}{!}{%
  \includegraphics{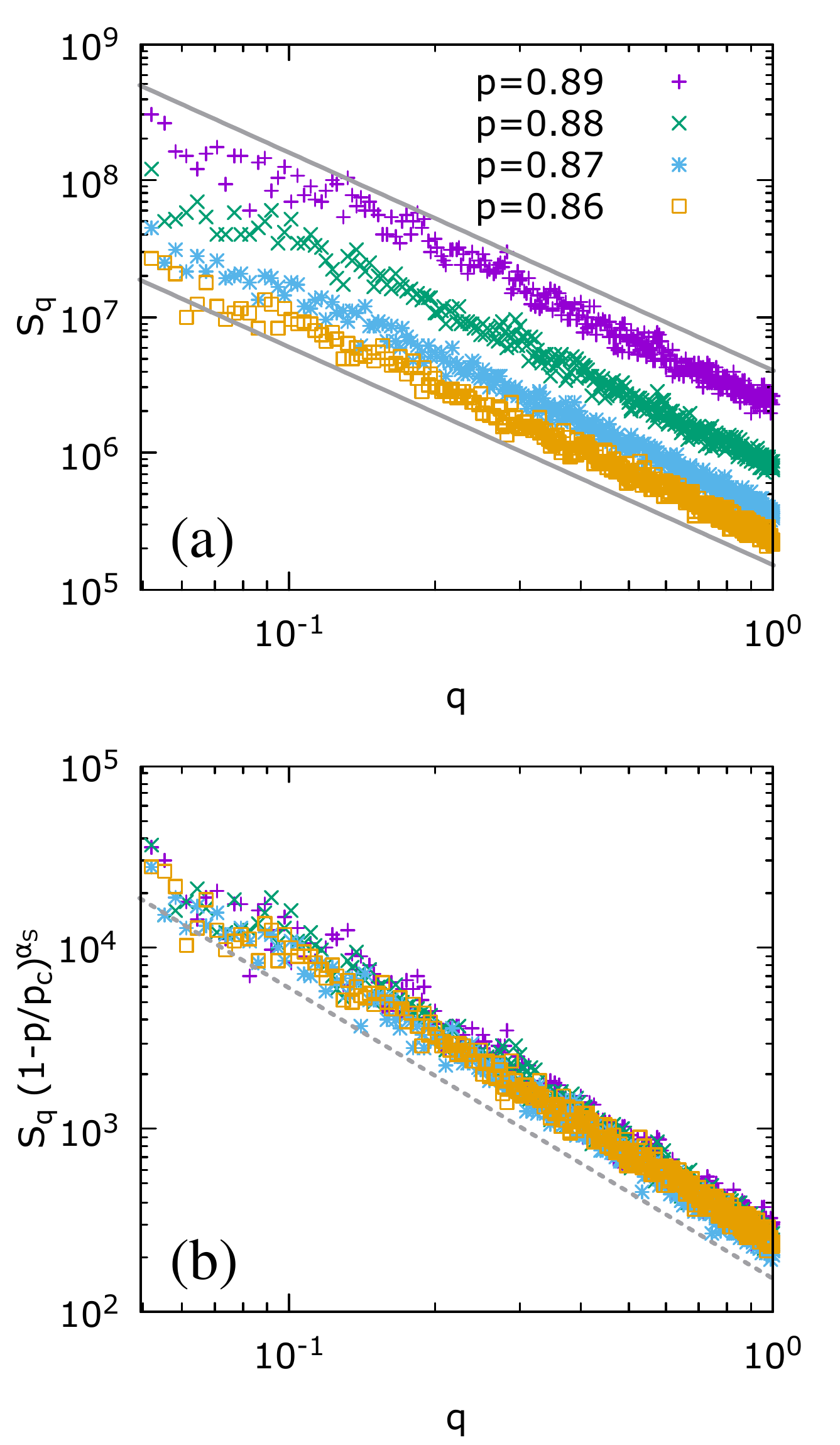}
}
\caption{Structure factor of the displacement field of the front for $p \lesssim p_c \approx 0.91$. (a) The front presents a self-affine structure, $S(q)\sim 1/q^{1+2\zeta}$ with $\zeta \approx 0.3$ (solid gray lines) but with a $p$-dependent prefactor, $(1-p/p_c)^{\alpha_S}$ with $\alpha_S \approx 2.4$, as evidenced by the rescaled plot in panel (b).}
\label{fig:sofq}       
\end{figure}

\begin{figure}
\resizebox{0.5\textwidth}{!}{%
  \includegraphics{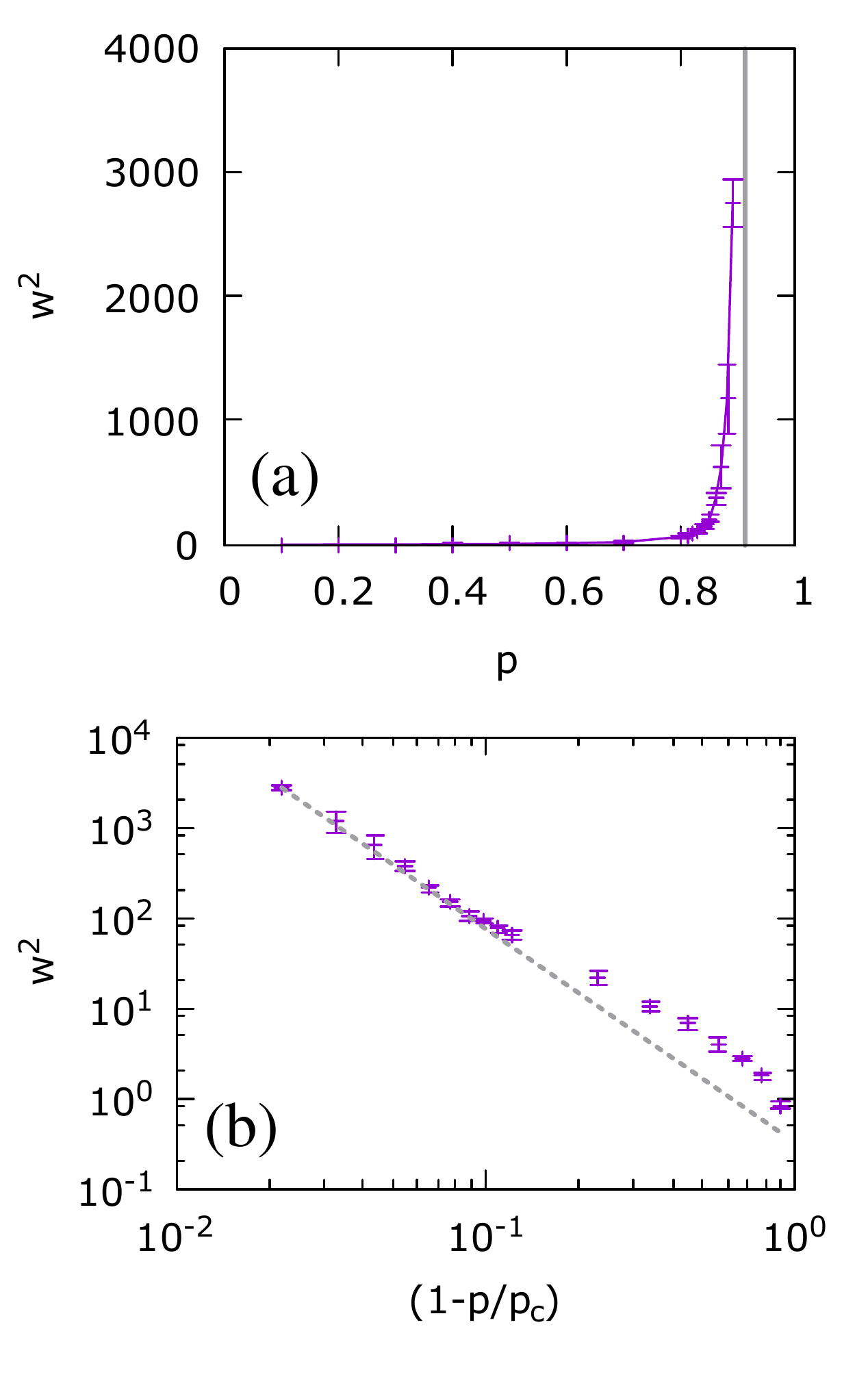}
}
\caption{ (a) Mean quadratic width $w^2$ (Eq.~\ref{eq:w2def}) for a finite front as a function of $p$. The width tend to diverge at the disorder threshold $p_c \approx 0.91$ (solid gray line). (b) Near $p_c$, $w^2 \sim (p-p_c)^{\alpha_w}$ with $\alpha_w \approx 2.4$ (dashed gray line).}
\label{fig:w2vsp}       
\end{figure}

We now study the geometrical properties of the front as a function of $p$, near the previously obtained $p_c$. 
In Fig.~\ref{fig:sofq}~(a) we show the structure factor $S(q)$ of the displacement field for various values of $p$ in the critical region. We find that it is particularly difficult to equilibrate the geometry of a large front near $p_c$ because its amplitude vanishes. We find however that for relatively short length-scales, the front develops a clear self-affine fractal structure, $S(q) \sim 1/q^{1+2\zeta}$, with roughness exponent $\zeta \approx 0.3 \pm 0.05$, and a $p$-dependent prefactor. Interestingly, the master curve of Fig.~\ref{fig:sofq}~(b) shows that the prefactor is critical, $S(q) \sim (1-p/p_c)^{\alpha_S}q^{1+2\zeta}$, with $\alpha_S\approx 2.37$. Since the mean quadratic width of the displacement field is $w^2 = \sum_q S(q) \approx \int_{2\pi/L}^{\pi} S(q)$, we get the scaling $w^2 \sim L^{2\zeta} (1-p/p_c)^{\alpha_w}$.
In Fig.~\ref{fig:w2vsp}~(a) we show that the predicted divergence of $w^2$ is present, and in Fig.~\ref{fig:w2vsp}~(b) we verify that $w^2 \sim (1-p/p_c)^{-\alpha_w}$ with the critical exponent $\alpha_w \approx 2.4$, indistinguishable from $\alpha_S$.

The non-trivial features of the critical behaviour near $p_c$, i.e. those that can not be explained by a naive homogenization approach discussed in Section \ref{sec:c}),  may be thus associated with these scale invariant geometrical properties.



\section{Conclusions}
\label{sec:conclusions}
Infection waves propagation in geographical landscapes is an old known phenomena (See, for instance~\cite{MurrayIIBook}). In this respect we note that Eqs.~\ref{eq:sir} have been used as a first basic model to understand the propagation of rabies in an initial population of susceptible (non rabid) foxes. Heterogeneity was also considered in a more realistic model, using a real map distribution of susceptibles. Such an approach is useful for a particular application of the model, but does not tell us about the universal or generic features that can arise, statistically, from disorder. Our work focus in that particular aspect.
We have studied the effect of random transmission heterogeneity in a diffusive SIR model with travelling waves solutions, by performing numerical simulations in an extended two dimensional system. We have found that the infection front changes its spatio-temporal fluctuating dynamics and its geometrical properties when increasing the fraction of sites ($p$) where local infection can not take place. 

In particular, propagation is completely arrested
at a non-trivial threshold value of the disorder ($p_c<1$). 
Moreover, approaching 
$p_c$ we have found a non-trivial critical behaviour for the speed $c$, amplitude $I_{max}$, shape $f_I$, 
structure factor $S(q)$ and mean quadratic width $w^2$ of the front, each one characterized by their critical exponents. 
Interestingly, the naive homogenization hypothesis  
(which consists in replacing $\beta$ in the 
homogeneous case by the spatial average in the heterogeneous case) 
describes qualitatively well the observed behaviour but is inaccurate to predict exponents and, in particular, underestimates the threshold $p_c$ for wave propagation. As a possible basic application, we can think $p$ as an heterogeneous local transmission rate due to a spatially random 
"vaccination" of susceptibles. Within this scenario, our results show that a naive homogenization hypothesis to account for the disorder, dangerously underestimates the level of vaccination needed to stop the infection. Besides this threshold issue, verifying the universality hypothesis of the critical exponents may open a way towards a quantitative characterization of the transport and geometry of infective fronts from a statistical physics approach, i.e. without relying too much in model details.

In some respects, the behaviour of the infective front propagation is reminiscent of the behaviour of elastic interfaces driven in a viscous random medium with a pinning landscape~\cite{Kardar1998,Fisher1998}. For instance, such kinds of models are successfully used to describe the propagation of domain walls in ferromagnets or the dynamics of contact lines of liquid on rough substrates. Indeed, in the the absence of disorder, an elastic interface becomes perfectly flat and propagates with a velocity proportional to the applied force $f$, as $c_0 \propto f$. 
In the presence of pinning the moving interface becomes spatially rough, temporally fluctuating and its velocity is reduced with respect to the free case. In particular, pinning yields a non-trivial critical value $f_c$ for the propagation of the interface, such that motion ceases for $f \leq f_c$. The velocity displays critical behaviour near the depinning threshold, $c \sim (f-f_c)^{\beta}$. Additionally the interface becomes self-affine at $f_c$, and $S(q)\sim 1/q^{1+2\zeta}$. These are all well known properties of the so-called depinning transition of elastic manifolds in random media. In all these respects, the behaviour of the infective front is qualitatively very similar to the one of a pinned elastic interface,  
if we think $S_c-S_0$ as an effective driving force for the displacement of the infective front. Moreover, 
the self-affine geometry of the front suggests  
that an effective elasticity of the front arises from 
the transverse diffusion of infectives.
There are important qualitative differences to note however. For a fixed size elastic string model, at depinning we find $w^2 \sim L^{2\zeta'}$ with no divergent prefactor in the limit $f \to f_c$, as it is observed for the front by making $p$ approach $p_c$ from below. This may be associated to the fact that the elastic interface do not change its internal structure as we approach $f_c$, only its displacement field changes, unlike the infective front which tends to deform in all directions and to disappear at $p_c$
\footnote{Forcing an analogy, one could still argue that the critical behaviour of the front is more similar to a magnetic field driven domain wall (DW) propagating in a ferromagnetic material near the order-disorder transition, since there the DW broadens and eventually disappears when the ordered ferromagnetic order is lost.}.
Apart from these qualitative similarities and differences, the roughness exponents for the best known depinning universality classes of driven elastic strings are clearly different to the one found for the infective front. This suggests that, from the general point of view of propagating self-affine interfaces~\cite{BarabasiBook}, infection fronts in the model described by Eqs.\ref{eq:sir} might belong to a new universality class. If so, are Eqs.~(\ref{eq:sir}) the minimal model for describing the new universality class?.

For many natural systems such as epidemics, forest fires or bacterial colony growth, the diffusive SIR model is a minimal model that allows to describe reaction-diffusion waves in an excitable media in general. The existence of critical behaviour in these kind of systems, suggests that some of the quantities we have obtained, such as the critical exponents, may be universal (at least whenever the real system displays an statistically uniform random heterogeneity in a reasonably extended region). Many of the properties we report here can be thus relevant for a basic understanding of the behaviour of more complex models describing more realistic situations, where spatial heterogeneity is known to be the rule rather than the exception.

\section*{Authors contributions}
All the authors were involved in the preparation of the manuscript.
All the authors have read and approved the final manuscript.

\section*{Acknowledgments}
We thank G. Abramson for useful discussions.
A.B.K and K.L are researchers of CONICET. K.L. acknowledges partial support from grant PIP/CONICET 2015 - 0100296 and PI/ UNRN 2017 40-B-552. A. B. K. acknowledges partial support from grants PICT2016-0069/FONCyT and UNCuyo C017, from Argentina.

\setcounter{section}{0}
\renewcommand{\thesection}{\Alph{section}}
\section{Appendix} 
\label{sec:numerics}
Equations~\ref{eq:sir} can be solved by the following implicit Euler scheme:
\begin{eqnarray}
I^{n+1}_{ij}&=&I^n_{ij}
+ \delta {\tilde t} \left[\tilde{\beta}_{ij} S^n_{ij}I^n_{ij} - \tilde{\gamma} I^n_{ij} \right] 
\nonumber \\
+ \delta{\tilde t} \;\tilde{D} &(&I^n_{i+1,j} + I^n_{i-1,j}
+ I^n_{i,j+1}+I^n_{i,j-1} 
-4 I^n_{ij}) \\
S^{n+1}_{ij} &=& S^{n}_{ij} - \delta {\tilde t}\; \tilde{\beta}_{ij} S^n_{ij} I^n_{ij}.
\end{eqnarray}
where the sub-indices denote the discretized two-dimensional space coordinates, and the super-index the discretized time variable.
By measuring time in units of $\beta^{-1}$ and space in units of $\sqrt{D\beta}$ we have the dimensionless parameters $\delta{\tilde t}=\beta \delta t$, and $\tilde{\gamma}=\gamma/\beta$, $\tilde{D}=D \beta /\delta x^2$. 
Disorder is implemented by assigning $\tilde{\beta}_{ij}=0$ with probability $p$ or $\tilde{\beta}_{ij}=1$ with probability $1-p$. 
The scheme is efficiently implemented using parallel computing in graphics processors, 
and results are obtained for grids as large as $2048 \times 2048$ sites.
In all our simulations we use ${\tilde \gamma}=0.2$, ${\tilde D}=1$ and $\delta {\tilde t}=0.01$.   
Movies from the simulations are provided as suplementary material.

\section{Appendix} 
\label{sec:analytics}
We review here some of the steps for obtaining the analytic steady-state solution of Eqs.~\ref{eq:sir} for a flat front in the homogeneous $p=0$ case (for more details see Ref.~\cite{MurrayIIBook}).

With a flat stripe initial condition $I(x,y,t=0)=I_0$ for $x_0< x < x_0+\delta x$ (such that $\partial_y I(x,y,t=0)=0$), $S(x,y,t=0)=S_0$, and periodic boundary conditions in the $y$-direction, we have $\partial_y I(x,y,t)=0$ for all $t>0$. By symmetry, the problem then becomes one dimensional:
\begin{eqnarray}
\dot{I}&=&\beta S I - \gamma I + D \partial_x^2 I 
\label{eq:dotI}
\\
\dot{S}&=&-\beta S I 
\label{eq:dotS}
\end{eqnarray}
We propose a steady-state (i.e. with no memory of the initial condition) wave solution $I=f(z)$, $S=g(z)$ with $z=x-ct$, with the boundary conditions $f(\pm \infty)=f'(\pm \infty)=0$,  $g(\infty)=S_0$, and $g(\infty)=S_1$,  the last one anticipating a residual number $S_1$ of remaining susceptibles after the wave passage. 
Then, 
Eq. \ref{eq:dotI} yields:
\begin{eqnarray}
-c f' &=& \beta f g - \gamma f + D f'' 
\label{eq:f}\\
-c g' &=& -\beta f g. 
\label{eq:g}
\end{eqnarray}
We linearize Eq. \ref{eq:f} near the leading edge, assuming $z \gg \lambda^{-1}$ (with $\lambda$ a characteristic distance self-consistently obtained below), where $g$ approaches the constant values $S_0$, to obtain 
\begin{equation}
-c f' = \beta S_0 f - \gamma f + D f'',
\end{equation}
which has a solution $f \sim e^{-\lambda z}$. Then, $\lambda$ satisfies: 
\begin{equation}
c\lambda = \beta S_0 - \gamma + D \lambda^2
\end{equation}
yielding the roots:
\begin{equation}
\lambda = \frac{c}{2D} \pm \sqrt{\left(\frac{c}{2D}\right)^2 - \frac{(\beta S_0 - \gamma)}{D}}.
\end{equation}
The travelling solution exists
if
\begin{equation}
c \geq 2\sqrt{D} \sqrt{\beta S_0 - \gamma},
\end{equation}
that is, only for $\beta S_0/\gamma>1$ or for $S_0>S_c$ with $S_c = \gamma/\beta$ the critical susceptible population. Equivalently, for a given $S_0$, we can write a critical transmission rate $\beta_c=\gamma/S_0$.  It can be shown that the actual velocity is the minimum of possible velocities \cite{MurrayIIBook}
\begin{equation}
c = 2\sqrt{D(\beta S_0 - \gamma)} = 
2 \sqrt{D\beta (S_0 - S_c)}
\end{equation}
The leading edge thus behaves as
\begin{equation}
f \sim \exp \left[-\frac{c}{2D} z \right]
\end{equation}
so there is a kind of ``Lorentz'' contraction of the front: the faster it travels the sharper is its leading edge.
A similar calculation applies for the trailing edge, $-z \ll \lambda^{-1}$ , where $g\sim S_1$ with $S_1$ the unknown remaining susceptibles
\begin{eqnarray}
f &\sim& \exp \left[\left(\frac{c}{2D} \mp \sqrt{\left(\frac{c}{2D}\right)^2 - \frac{(\beta S_1 - \gamma)}{D}}\right)z \right] \\ 
&=& \exp \left[\left(\frac{c}{2D} \mp \sqrt{\frac{\beta}{D}(S_0-S_1)}\right)z \right] 
\end{eqnarray}
The asymmetry of the shape seen in numerical simulations 
(see Fig.\ref{fig:shape})
shows that we must take the minus sign (the trailing edge is less sharp than the leading edge).
To obtain $S_1$ we use that since $g>0$, $f=cg'/g$ from Eq.~\ref{eq:g}, and replace $f$ in Eq.~\ref{eq:f}, obtaining:
\begin{equation}
    Df''+cf'+cg'-\frac{c\gamma}{\beta} \frac{d}{dz} \log g =0.
\end{equation}
Integration over $z$ thus gives:
\begin{equation}
    Df'+cf+cg-\frac{c\gamma}{\beta} \ln g = \text{cte}.
\end{equation}
Evaluating this expression in $z=\pm \infty$ using the assumed boundary conditions yields a transcendental equation for $S_1$:
\begin{equation}
{\frac{S_1}{S_0} - 1 }= \frac{S_c}{S_0}
\ln \frac{S_1}{S_0}
\end{equation}
implying $0< S_1 < S_c < S_0$. 
This shows in particular that the infected in the trailing edge can not trigger a wave going backwards, because $S_1<S_c$. We also note that $S_1$ is independent of $D$.

\bibliographystyle{ieeetr}
\bibliography{SIRDisordered.bib}

\end{document}